\documentclass[british,a4paper]{amsart}
\usepackage{iskra}
\hypersetup{
pdfstartview={FitH},
pdftitle={Trace Relations in Deformed Gauge Theories},
pdfauthor={Madhusudhan Raman, Aditi Shahani},
}

\usepackage{cleveref}

\title{Trace Relations in Deformed Gauge Theories}

\author[M.~Raman]{Madhusudhan Raman}
\address{M.~Raman: Department of Physics and Astrophysics\\
  University of Delhi, Delhi 110 007, India}
\email{mraman@physics.du.ac.in}

\author[A.~Shahani]{Aditi Shahani}
\address{A.~Shahani: Homer L. Dodge Department of Physics and Astronomy\\
  University of Oklahoma, 440 W. Brooks St. Norman, OK 73019}
\email{aditi.j.shahani-1@ou.edu}

\date{\DTMnow}

\begin{document}

\begin{abstract}
  We compute trace relations governing chiral ring elements of fully
  $\Omega$-deformed $\mathcal{N}=2^{\star}$ gauge theories with
  $\mathrm{SU}(N)$ gauge groups by demanding the regularity of the
  fundamental $qq$-character.
\end{abstract}

\maketitle
\tableofcontents

\section{Introduction}
\label{sec:intro}

Consider an $ N \times N $ matrix $ \Phi $, whose eigenvalues are the
roots of its characteristic polynomial $P_{N}(x )$, defined in
the usual way as
\begin{equation}
    P_{N}(x) = \operatorname{det} \left(x - \Phi \right) \ .
\end{equation}
An elementary result in linear algebra due to Cayley
and Hamilton establishes that any such matrix satisfies its own
characteristic equation: $ P _{N}(\Phi ) = 0 $.

It is an instructive exercise to construct this characteristic
polynomial recursively using an algorithm due to Faddeev and
LeVerrier, which proceeds as follows. Starting with $P_{0}(\Phi) = 1$,
one builds up to $P_{N}(\Phi)$ using
\begin{equation}
    P_{k}(\Phi) = \Phi \, P_{k-1}(\Phi) -\frac{1}{k} \operatorname{Tr}
    \left[\Phi\,P_{k-1}(\Phi)\right] \ .
\end{equation}
In this way, one finds that the equation
$\operatorname{Tr} P_{N}(\Phi) = 0 $ expresses
$\operatorname{det} \Phi$ in terms of traces of powers of $\Phi$. By
the same token, one also finds that higher traces of the form
$\operatorname{Tr} \Phi^{N+k}$ with $k\geq 1$ can be expressed as a
linear combination of products of traces of lower powers of
$\Phi$. For example, using the Faddeev-LeVerrier algorithm it is easy
to show that a generic traceless $ 3 \times 3 $ matrix satisfies
\begin{equation}
\operatorname{Tr} \Phi ^{4} = \frac{1}{2} \left[ \operatorname{Tr} \Phi ^{2}
\right] ^{2} \ .
\end{equation}
Relations of this kind are called trace relations, and will be the
object of our study.

Since the above discussion is true for any finite-dimensional matrix,
we might consider the case where $ \Phi $ specifies the vacuum
expectation value on the Coulomb branch of the adjoint scalar in the
$ \mathcal{N}=2 $ vector multiplet. Traces of powers of this operator
are chiral, i.e.~they are annihilated by all supercharges of one
chirality. Further, the classical moduli space of these theories
receives nonperturbative corrections --- see
\cite{Intriligator:1995au} for a review --- and as a consequence the
classical trace relations are modified too. As an illustrative
example, employing the notation
$ u \equiv \left\langle \operatorname{Tr} \Phi ^{2} \right\rangle $
and $ q = \Lambda ^{4} $, the instanton counting parameter, the trace
relations in the pure $ \mathrm{SU}(2) $ gauge theory are
\cite{Cachazo:2002ry}:
\begin{equation}
\begin{aligned}
  \left\langle \operatorname{Tr} \Phi ^{4} \right\rangle &= \frac{1}{2} u ^{2} \mathcolor{WildStrawberry}{+ 4q} \ , \\
  \left\langle \operatorname{Tr} \Phi ^{6} \right\rangle  &= \frac{1}{4} u ^{3} \mathcolor{WildStrawberry}{+ 6 q u} \ , \\
  \left\langle \operatorname{Tr} \Phi ^{8} \right\rangle &= \frac{1}{8} u ^{4} \mathcolor{WildStrawberry}{+ 6 q u ^{2} + 12 q ^{2}} \ ,
\end{aligned}
\end{equation}
and so on. The highlighted terms in the above equation correspond to
nonperturbative quantum corrections to the classical trace
relations. We are interested in these kinds of quantum corrections,
but in the presence of gravitational couplings as well.

The class of theories we will focus on are massive deformations of
$ \mathcal{N}=4 $ super Yang-Mills theories with gauge group
$ \mathrm{SU}(N) $, where one of the three chiral multiplets is given
a mass $ m $. The non-trivial physics of these theories is in the
instanton sector, which as we have seen above corrects the Coulomb
moduli space \cite{Seiberg:1994rs,Seiberg:1994aj}. The \emph{exact}
partition function --- which determines, among other things, the
low-energy effective action --- can be computed using equivariant
localisation \cite{Nekrasov:2002qd,Nekrasov:2003rj}, and this in turn
necessitates the introduction of the $ \Omega $-background: a
two-parameter supergravity background that breaks four-dimensional
Poincar\'e invariance but preserves a part of the (deformed)
supersymmetry.

Now, while the partition function of the $ \Omega $-deformed gauge
theory, in the limit of vanishing deformation parameters, is able to
reproduce the prepotential of the $ \mathcal{N}=2 $ gauge theory, the
partially or even fully deformed gauge theory partition function is a
natural object of study for many reasons. When expanded in the limit
of small deformations, higher-order terms compute gravitational
couplings of the gauge theory
\cite{Manschot:2019pog,John:2022yql,Ashok:2023tsm}.  Even partial
deformations are interesting: the Nekrasov-Shatashvili limit
\cite{Nekrasov:2009rc} in which only one of the two
$ \Omega $-deformation parameters is turned on leads to a quantisation
\cite{Nekrasov:2011bc,Nekrasov:2013xda} of the classical integrable
system underlying the \emph{undeformed} gauge theory
\cite{DHoker:1999yni}. Most recently, this limit was used to shed
light on the exact spectrum of a non-local deformation of quantum
mechanics \cite{Grassi:2018bci}.

A more striking development is the AGT-W correspondence
\cite{Alday:2009aq,Gaiotto:2009we,Wyllard:2009hg}, an avatar of which
provides a precise mapping between the deformed instanton partition
function of an $ \mathcal{N}=2 $ gauge theory with rank-$ 1 $ (higher
rank) gauge group and the Virasoro (higher spin) conformal blocks of a
two-dimensional conformal field theory. Under this correspondence, the
deformed instanton partition functions of the theories we study in
this note map to the torus $ 1 $-point Virasoro/$ \mathcal{W}_{N} $
block \cite{Poghossian:2009mk,Fateev:2009aw,Das:2020fhs}. It is
expected that chiral operators in the gauge theory map to the infinite
family of conserved integrals of motion associated to a
two-dimensional conformal field theory, the so-called KdV charges
\cite{Nekrasov:2009rc,Alba:2010qc,Fateev:2011hq}. This makes the task
of determining the deformed chiral ring relations interesting, for
although such relations do not by themselves fix this map, they would
supply a strong consistency check on any proposal for the same.

The $\Omega$-deformed trace relations have been the subject of some
interest in the past. In \cite{Fucito:2015ofa}, the deformed trace
relations for gauge theories with fundamental matter and their
conformal field theory duals were studied, and in
\cite{Beccaria:2017rfz}, trace relations in the chiral ring of
$ \Omega $-deformed $ \mathcal{N}=2 ^{\star } $ theories were
studied. This latter work was based on multi-instanton computations,
and certain conjectures for the deformed chiral ring relations were
presented. The methodology employed here was to compute
$ k $-instanton corrections to chiral ring basis elements
$ \mathcal{O}_{n} $ in the $ \Omega $-deformed gauge theories (with
and without matter), after which the chiral ring elements
$ \mathcal{O}_{n>N} $ were expressed in terms of the
$ \mathcal{O}_{n \leq N} $ by rather laborious sequence of series
inversions.

These are no doubt impressive and difficult computations. Having said
that, there are obvious difficulties associated with carrying out such
multi-instanton computations and inversions, especially for the case
of higher-rank gauge groups. (For $ \mathrm{SU}(N) $ gauge theories,
this would involve $ N-1 $ series inversions to express the
independent Coulomb vevs in terms of the gauge invariant Coulomb
moduli and their derivatives!) It is therefore desirable to explore
the use of alternative methods of arriving at these conclusions. In
this brief note, we will verify these conjectures by requiring that
certain observables called $ qq $-characters are regular, following in
the footsteps of \cite{Jeong:2019fgx}. In
\Cref{sec:trace-relat-deform} we explain this methodology and
introduce the $ qq $-character for the $ \mathcal{N}=2 ^{\star } $
theory. In \Cref{sec:results}, we write down the deformed trace
relations in gauge theories with low rank to demonstrate the validity
and relative ease with which earlier results may be
recovered. (\Cref{sec:fourier-expansions} gives expressions for the
Fourier series employed in the resummations we discuss present in this
section.) We conclude in \Cref{sec:discussion} with a brief discussion
of some possible directions for future work.

\subsection*{Acknowledgements}
The authors are grateful to Sujay Ashok for discussions. MR is
supported by an Inspire Faculty Fellowship.

\section{Trace Relations via Deformed Characters}
\label{sec:trace-relat-deform}
In the context of supersymmetric gauge theories on flat space, chiral
operators are local operators that are annihilated by all the
supercharges of a particular chirality. From the nilpotence of the
supercharge it follows that such operators can only be defined up to
equivalence. Finally, it follows from the supersymmetry algebra ---
here it is important that we are in flat spacetime --- that products
of chiral operators are chiral, and that the expectation values of
such operators is independent of their spacetime position. This
equivalence class of such operators is called the chiral ring, and has
the structure of a finitely generated commutative ring
\cite{Novikov:1983ee,Cachazo:2002ry}.

To define the trace relations, which follow from the fact that the
chiral ring is finitely generated, we must first define a basis for
our chiral ring, which we choose in standard fashion to be operators
of the form
\begin{equation}
\mathcal{O}_{n} = \operatorname{Tr} \Phi ^{n} \ ,
\end{equation}
for $ n \in \left\{1, \cdots , N\right\} $.  The field
$ \mathcal{\Phi } $ is the adjoint scalar in the $ \mathcal{N}=2 $
vector multiplet, and since we will focus our attention on
$\mathrm{SU}(N)$ gauge theories, we will impose the condition
$\operatorname{Tr} \Phi = 0 $. Any product of the operators
$ \mathcal{O}_{n} $ is an element of the chiral ring, and such
products shall henceforth be denoted
\begin{equation}
\mathcal{O}_{\mathbf{n} } = \prod _{k=1} ^{\left\vert \mathbf{n}  \right\vert  } \mathcal{O}_{n _{k}} \ .
\end{equation}
Boldface quantities such as $ \mathbf{n} $ shall henceforth be used to
refer to the tuples
$ \left\lbrace n _{1}, n _{2}, \cdots \right\rbrace $ with cardinality
$ \left\vert \mathbf{n} \right\vert $. Correlation functions of chiral
operators will be denoted
\begin{equation}
    u_{\mathbf{n}} = \left\langle \mathcal{O}_{\mathbf{n}} \right\rangle \ .
\end{equation}
In flat spacetime, from cluster decomposition it follows that
\begin{equation}
	u_{n_{1},n_{2},\cdots} = u_{n_{1}} u_{n_{2}} \cdots \ ,
\end{equation}
i.e.~correlation functions of chiral operators factorise. This is not
true on curved spacetimes, and as we will see quite explicitly in the
next section.

In the introduction to this note, we saw the Faddeev-LeVerrier
algorithm, a simple recursive method that generates chiral ring
relations. It is useful here to introduce another presentation of the
same, which expands out the characteristic polynomial associated to
$ \Phi $ in the following way:
\begin{equation}
  \label{eq:classical-trace-relations}
\begin{aligned}
  \operatorname{det} \left( x-\Phi  \right) &= x ^{N} \operatorname{exp} \operatorname{Tr} \log \left( 1-\frac{\Phi }{x}  \right) \ , \\
  &= x ^{N} \operatorname{exp} \left\lbrace - \sum_{n=1}^{\infty } \frac{1}{n x ^{n}} \operatorname{Tr} \Phi ^{n} \right\rbrace \ , \\
  &= x ^{N} - \frac{x ^{N-2}}{2}  \mathcal{O}_{2} - \frac{x ^{N-3}}{3}  \mathcal{O}_{3} - \frac{x ^{N-4}}{4}  \left(\mathcal{O}_{4} - \frac{1}{2} \mathcal{O}_{2,2} \right) + \cdots \ .
\end{aligned}
\end{equation}
By definition, this is a degree $ N $ polynomial in the classical
vacuum expectation values of the scalar $ \Phi $, and so the
coefficients of all the negative powers of $ x $ in the above
expansion must be zero. This gives us the classical chiral ring
relations. For gauge theories on an $\Omega$-deformed
$\mathbb{R}^{4}$, in addition to nonperturbative corrections of the
kind we saw in \Cref{sec:intro}, the chiral ring relations are
modified by gravitational corrections that depend on the choice of
$\Omega$ background.  The task of determining the deformed chiral ring
relations is then the task of quantifying in a nonperturbatively exact
manner what these corrections are.

\subsection{Instantons and $ \Omega  $-Deformation}
The class of theories we will focus on are $\mathcal{N}=2^{\star}$
supersymmetric gauge theories, i.e.~massive deformations of
$ \mathcal{N}=4 $ super Yang-Mills theories with gauge group
$ \mathrm{SU}(N) $ where one of the three chiral multiplets is given a
mass $ m $. We will consider this theory on the Coulomb branch, where
the adjoint scalar $ \Phi $ acquires a vacuum expectation value
$\Phi = \operatorname{diag}(a_{1}, \cdots, a_{N})$ subject to the
condition $ \operatorname{Tr} \Phi =0 $. These theories are conformal,
enjoy S-duality symmetry, and their perturbative dynamics is
$ 1 $-loop exact, so the only non-trivial physics of these theories
resides in the instanton sector, which was first completely solved for
in \cite{Donagi:1995cf,DHoker:1997hut}. The quantum-corrected moduli
space of the supersymmetric gauge theory is, in this picture,
identified with the moduli space of an elliptic curve. Soon after, the
exact partition function was computed using equivariant localisation
\cite{Nekrasov:2002qd,Nekrasov:2003rj}, thereby supplying a rigorous
derivation of the Seiberg-Witten solution.

The localisation procedure requires that the integrals over
$ k $-instanton moduli spaces be regularised
\cite{Flume:2002az,Bruzzo:2002xf}, and this is done by working in the
so-called $ \Omega $-background, a supergravity background that breaks
Poincar\'e invariance and which is specified by two parameters
$ \epsilon _{1} $ and $ \epsilon _{2} $. It is helpful to introduce
the following shorthand for the sum $(s)$ and product $(p)$ of the
$\Omega$-deformation parameters:
\begin{equation}
    s = \epsilon_{1}+\epsilon_{2} \quad \mathrm{and} \quad p = \epsilon_{1}
    \epsilon_{2} \ .
\end{equation}

It will be useful to recall the manner in which the deformed partition
function of $ \mathcal{N}=2 ^{\star } $ gauge theories with gauge
group $ \mathrm{SU}(N) $ is constructed. Consider an $ N $-vector of
Young diagrams $ \bm{Y} = \left( Y _{1}, \cdots , Y _{N} \right) $
with $ \left\vert \bm{Y} \right\vert $ boxes. The deformed instanton
partition function of the gauge theory can be expressed as the
following statistical sum over all such vectors of Young diagrams:
\begin{equation}
  \label{eq:instanton-partition-function}
\mathcal{Z} \equiv \mathcal{Z}\left( a _{u}  , m , q; \epsilon _{1}, \epsilon _{2} \right) = \sum_{\bm{Y} }^{} q ^{\left\vert \bm{Y}  \right\vert } \mu _{\bm{Y} } \ ,
\end{equation}
where $ m $ is the mass of the adjoint hypermultiplet,
$ q = e^{2 \pi i \tau } $ is the instanton counting parameter, and
$ \tau $ the complexified gauge coupling. Correlation functions are
also expressed as statistical sums weighted by this measure factor
$ \mu _{\bm{Y} } $:
\begin{equation}
  \label{eq:chiral-correlator-general}
  \left\langle \mathcal{O} \right\rangle  = \frac{1}{\mathcal{Z}} \sum_{\bm{Y} }^{} q ^{\left\vert \bm{Y}  \right\vert } \, \mu _{\bm{Y} } \, \mathcal{O}_{\bm{Y} } \ .
\end{equation}
We refer the reader to \cite{Billo:2012st} for a more comprehensive
discussion of the form taken by the weight $ \mu _{\bm{Y} } $ and
especially the operator $ \mathcal{O}_{\bm{Y} } $ appropriate to
$ \mathcal{O} = \operatorname{Tr} e^{z \Phi } $, the generating
function for chiral operators. Since these explicit localisation
computations were performed in \cite{Beccaria:2017rfz}, we do not
repeat them in this paper.

\subsection{Doubly Deformed Characters}
We review now briefly the method introduced by \cite{Jeong:2019fgx}
for studying the trace relations in $ \Omega $-deformed gauge
theories. We also discuss the $ qq $-character of the
$ \mathcal{N}=2 $ theory with adjoint matter, which has received less
attention than its pure or fundamental counterparts.

The $ qq $-characters were first defined in
\cite{Nekrasov:2015wsu,Nekrasov:2016qym,Nekrasov:2016ydq} as composite
operators of the so-called $ \mathcal{Y} $-observable
\begin{equation}
  \label{eq:Y-observable}
\mathcal{Y}(x) = x ^{N} \operatorname{exp} \left\lbrace - \sum_{n=1}^{\infty } \frac{1}{n x ^{n}} \operatorname{Tr} \Phi ^{n} \right\rbrace \ ,
\end{equation}
that, in the classical limit, reduces to
\cref{eq:classical-trace-relations}, the characteristic
polynomial. When evaluated on a specific instanton configuration
\cite{Losev:2003py} of the kind contributing to the sum in
\cref{eq:instanton-partition-function} we have
\begin{equation}
  \label{eq:y-observable-intermediate}
\left[ \mathcal{Y}(x) \right]_{\bm{Y} } = \prod _{u=1} ^{N} \left[ \left( x-a _{u} \right) \prod _{\smol{\yng(1)} \in Y _{u}} ^{} \frac{\left( x - a _{u}- c _{\smol{\yng(1)} }- \epsilon _{1} \right)\left( x - a _{u}- c _{\smol{\yng(1)} }- \epsilon _{2} \right)}{\left( x - a _{u}- c _{\smol{\yng(1)} } \right)\left( x - a _{u}- c _{\smol{\yng(1)} }- s  \right)}  \right] \ ,
\end{equation}
where in the above equation, the product over
$ \smol{\yng(1)} \in Y _{u} $ denotes a product over all boxes
labelled by coordinates $ (i,j) $ in the Young diagram $ Y _{u} $ and
\begin{equation}
c _{\smol{\yng(1)}} = \epsilon _{1} \left( i-1 \right) + \epsilon _{2}\left( j-1 \right) \ .
\end{equation}
It is easy to see that \cref{eq:y-observable-intermediate} will have
many pairwise cancellations, which in the end lead to the following
simplified expression over the ``edge'' of the Young diagram:
\begin{equation}
\left[ \mathcal{Y}(x) \right]_{\bm{Y } } = \prod _{u=1} ^{N}
\frac{\prod _{\smol{\yng(2,2)} \in \partial _{+}Y _{u}} ^{} \left( x-a _{u}- c _{\smol{\yng(2,2)}} \right)}{\prod _{\smol{\yng(2)} \in \partial _{-}Y _{u}} ^{} \left( x-a _{u}- c _{\smol{\yng(2)}} \right)} \ .
\end{equation}
In the above equation, the products over $ \partial _{\pm }Y _{u} $
are over the boxes of Young diagrams that have one box more $ (+) $ or
less $ (-) $ than any reference diagram
$ \smol{\yng(2,1)} \in Y _{u} $.

While the $ \mathcal{Y} $-observable has singularities, a composite
operator $ \mathcal{X}(x) $ can be built out of it that, by
construction, is necessarily polynomial and therefore regular in the
independent variable $ x $. Our focus in this paper will be on the
fundamental $ qq $-character of the $ \mathcal{N}=2 ^{\star } $ theory
with gauge group $ \mathrm{SU}(N) $
\cite[eq.~(7.6)]{Nekrasov:2015wsu}, which is given in terms of an
\emph{infinite} sum over Young diagrams or, what is the same thing,
integer partitions. Let $ \lambda  $ be a Young diagram:
\begin{equation}
\lambda = \left( \lambda _{1}, \lambda _{2}, \cdots \right) \quad \mathrm{with} \quad \left\vert \lambda \right\vert = \lambda _{1} + \lambda _{2} + \cdots \ .
\end{equation}
The fundamental $ qq $-character $ \mathcal{X}(x) $ is given by:
\begin{equation}
  \label{eq:fund-qq-character-1}
\begin{aligned}
  \mathcal{X}(x) &= \mathcal{Y}(x+s ) \sum_{\lambda }^{} q ^{\left\vert
  \lambda  \right\vert } f _{\lambda } \ , \\
  &= \mathcal{Y}(x+s ) \left[ 1+ q \, f _{\smol{\yng(1)} } + q ^{2} \left( f
  _{\smol{\yng(2)}} + f _{\smol{\yng(1,1)}}\right) + q ^{3} \left( f
  _{\smol{\yng(3)}} + f _{\smol{\yng(2,1)}}+ f _{\smol{\yng(1,1,1)}}\right)+
  \cdots  \right]
\end{aligned}
\end{equation}
where
\begin{equation}
  \label{eq:fund-qq-character-2}
f _{\lambda } = \prod _{\smol{\yng(1)} \in \lambda } ^{}  S \left( m h
_{\smol{\yng(1)} } + s l _{\smol{\yng(1)} } \right) \frac{\mathcal{Y}\left(x
+ \sigma _{\smol{\yng(1)} }-m\right) \mathcal{Y}\left( x + \sigma
_{\smol{\yng(1)} }+m + s  \right)}{\mathcal{Y}\left(x + \sigma
_{\smol{\yng(1)} }\right) \mathcal{Y}\left( x + \sigma _{\smol{\yng(1)} }+
s  \right)} \ .
\end{equation}

In the above equations, the sum is over all Young diagrams $ \lambda $
and the product is over all boxes $ \smol{\yng(1) } $ in the Young
diagram $ \lambda $, with $ h _{\smol{\yng(1)} } $ and
$ l _{\smol{\yng(1)} } $ the hook- and leg-lengths corresponding to
$ \smol{\yng(1)} $ respectively. Finally,
\begin{equation}
\sigma _{\smol{\yng(1)} } = m \left( i-j \right) + s \left( 1-j \right) \ ,
\end{equation}
where $ \smol{\yng(1)} $ has coordinates $ (i,j) $, and
\begin{equation}
  \label{eq:s-function-def}
S(x) = 1+\frac{p}{x(x+s )} \ .
\end{equation}
For example, up to $ 2 $-instantons, the expression for the
fundamental character is
\begin{equation}
  \begin{aligned}
  \mathcal{X}(x) &= \mathcal{Y}(x+s ) + q \, S(m) \,
  \frac{\mathcal{Y}(x-m)\mathcal{Y}(x+m+s )}{\mathcal{Y}(x)} \\
  &\quad + q ^{2} \left( S(2m+s ) S(m)
  \frac{\mathcal{Y}(x-m)\mathcal{Y}(x+2m+s )}{\mathcal{Y}(x+m)} \right. \\
  &\quad \quad \quad \quad \left. + S(m)S(2m) \frac{\mathcal{Y}(x+m+s
  )\mathcal{Y}(x-2m-s )}{\mathcal{Y}(x-m-s )}  \right) + \cdots \ .
  \end{aligned}
\end{equation}
The $ qq $-characters generalise the $ q $-characters of
\cite{Nekrasov:2013xda,Fucito:2011pn,Fucito:2012xc}, and reduce to
them in the Nekrasov-Shatashvili limit, where $ p \rightarrow 0 $
while $ s $ is held fixed.

The regularity of the fundamental $ qq $-character was first used by
\cite{Jeong:2019fgx} to provide a simple framework within which the
(deformed) chiral ring of an $ \mathcal{N}=2 $ gauge theory can be
investigated. Their methodology was simple: the regularity of the
$ qq $-character meant that an expansion of
$ \left\langle \mathcal{X}(x) \right\rangle $ around $ x = \infty $
should yield a polynomial, and therefore that the coefficients
of $ x ^{-n} $ for $ n \in \mathbb{N} $ should vanish identically,
giving the suitably deformed trace relations. Using this unifying
picture, they were able to independently verify some results for pure
gauge theories \cite{Beccaria:2017rfz} and theories with fundamental
matter \cite{Fucito:2015ofa}. In particular, the fact that the
commutative ring structure is replaced by a differential ring
naturally follows from the simple observation that
\begin{equation}
  \label{eq:u2-shift}
\left\langle \mathcal{O}_{2} ^{n} \mathcal{O} \right\rangle = \left( u _{2} - \frac{p}{2 \pi i} \frac{\mathrm{d} }{\mathrm{d} \tau }  \right)^{n} \left\langle \mathcal{O} \right\rangle \ ,
\end{equation}
a fact that is easily verified using
\cref{eq:chiral-correlator-general} and which we will use
repeatedly. Our goal in the following section will be to extend this
work to theories with adjoint matter, and we will proceed by first
expanding out the $ qq $-character in \cref{eq:fund-qq-character-1} at
large-$ x $ after using \cref{eq:Y-observable}, then collecting the
coefficients of poles in $ x $ and demanding that they
vanish. Whenever we have correlation functions of the form
$ \left\langle \mathcal{O}_{2}^{n} \mathcal{O} \right\rangle $, we use
\cref{eq:u2-shift}. Finally, since we are working order-by-order in an
instanton expansion, we will need to resum our results into
appropriate combinations of the Fourier series defined in
\Cref{sec:fourier-expansions}.

\section{Results}
\label{sec:results}
We now present the results of the computations outlined in the
previous section. In all the examples we have explored, we are able to
confirm --- independently and with significantly less effort --- the
results of earlier computations, especially those of
\cite{Beccaria:2017rfz}. In order to facilitate this comparison, we
employ a shift
\begin{equation}
    m \rightarrow m - \frac{s}{2} \ ,
\end{equation}
in the rest of this paper. We also use the notation
\begin{equation}
\mathcal{C} = 4 \left( m ^{2} + p \right) - s ^{2} \ ,
\end{equation}
and the shorthand
\begin{equation}
X' = \frac{1}{2 \pi i} \frac{\mathrm{d} }{\mathrm{d} \tau } X \ ,
\end{equation}
for any function $ X \equiv X(\tau ) $.

A word on resummation: our strategy has been to organise the
gravitational corrections to the trace relations to
$ \left\langle \mathcal{O}_{k} \right\rangle $ by resumming the
Fourier expansions accompanying all possible terms of the form
$ \mathcal{C}^{a}p ^{b} s ^{c} $ such that $ k = 2(a+b)+c $. The
resummations presented in the following sections are uniformly tested
to \emph{at least} $ 6 $-instantons, and this proves to be more an
enough for chiral correlators of low enough mass dimension. However,
in order to resum certain coefficients, we have needed to go up to
$ 10 $-instantons. We will have more to say about this after we have
had the opportunity to present some results.

\subsection*{\boldmath $\mathrm{U}(1)$}
All the computations outlined in the previous section are formally
valid for the case of $N=1$. We find by explicit computation and
resummation that
\begin{equation}
  \label{eq:u1-o2}
        \left\langle \mathcal{O}_{2} \right\rangle = + \,
        \frac{1}{48}\, \mathcal{C}\left(E_{2} - 1\right) \ ,
\end{equation}
\begin{equation}
\left\langle \mathcal{O}_{3} \right\rangle = - \,
        \frac{3}{4} \, \mathcal{C}s \, E_{3} \ ,
\end{equation}
\begin{equation}
    \begin{aligned}
        \left\langle \mathcal{O}_{4} \right\rangle &=
        \mathcolor{Black}{\frac{\mathcal{C}p}{1440} \left( 5 E _{2}^{2} + E
        _{4} - 6 \right)} - \frac{\mathcal{C}s ^{2}}{240} \left( E _{4}-1
        \right) \\
        &\quad + \frac{\mathcal{C}^{2}}{11520} \left( 10 E _{2}^{2}-E _{4}-30 E _{2}+21 \right) \ ,
    \end{aligned}
\end{equation}
\begin{equation}
\begin{aligned}
  \left\langle \mathcal{O}_{5} \right\rangle &= \frac{5 \mathcal{C}ps}{2}
  \left( E _{5} - E _{3}' \right) - \frac{5 \mathcal{C}s ^{3}}{4} E _{5} \\
  &\quad -\frac{5 \mathcal{C}^{2}s}{32} \left( 2 E _{5}-2 E _{3}' + E _{2} E
  _{3} - E _{3} \right) \ ,
\end{aligned}
\end{equation}
and so on. These results are derived by demanding regularity of the
$ qq $-character and are in good agreement with earlier
computations \cite[App.~F.2]{Beccaria:2017rfz}. 

As another consistency check, it is possible to arrive at results for
$ \mathcal{N}=1 ^{\star } $ theories starting from results pertaining
to $ \mathcal{N}=2 ^{\star } $ theories by tuning the vevs of the
adjoint scalar to the locations of $ \mathcal{N}=1 $ supersymmetric
vacua. (Recall that the $ \mathcal{N}=1 ^{\star } $ theory is a
massive deformation of $ \mathcal{N}=4 $ theory, arrived at by giving a
mass $ m $ to all three chiral multiplets.) This procedure is trivial
in the case of a theory with $ \mathrm{U}(1) $ gauge group since the
theory has only one vacuum. We can compare \cref{eq:u1-o2} with
\cite[Sec.~4.1]{Fucito:2005wc}, where exact gravitational corrections
to this correlator are computed in the special case
$ s = 0 \Rightarrow \epsilon _{1} = - \epsilon _{2} = \hbar $. Taking
this limit in our result above, we find
\begin{equation}
\left\langle \mathcal{O}_{2} \right\rangle _{\mathcal{N}=1 ^{\star }} =
\left( m ^{2} - \hbar ^{2} \right) \textcolor{Black}{\left(
\frac{E_{2}-1}{12}  \right)} \ ,
\end{equation}
which agrees with \cite{Fucito:2005wc} after a rescaling
$ (m _{\mathrm{here}},\hbar _{\mathrm{here}}) \rightarrow 2 ^{-1/2} (m
_{\mathrm{there}}, \hbar _{\mathrm{there}} ) $.

\subsection*{\boldmath $\mathrm{SU}(2)$}
Requiring that the $ qq $-character is regular implies the following
trace relations:
\begin{equation}
        \left\langle \mathcal{O}_{3} \right\rangle = - \,
        \frac{3}{2} \, \mathcal{C}s \, E_{3} \ ,
\end{equation}
\begin{equation}
  \label{eq:su2-o4}
    \begin{aligned}
\left\langle \mathcal{O}_{4} \right\rangle &= \frac{u ^{2}}{2} - p u' + \frac{\mathcal{C}}{12} \left( E _{2}-1 \right) u \\
        &\quad + \frac{\mathcal{C}p}{480} \left( 5 E _{2}^{2} - E _{4} - 4 \right) - \frac{\mathcal{C}s ^{2}}{120} \left( E _{4}-1 \right) \\
        &\quad \quad - \frac{\mathcal{C}^{2}}{1440} \left( 5 E _{2}^{2}-E _{4}-5 E _{2}+1 \right) \ ,
    \end{aligned}
  \end{equation}
\begin{equation}
\begin{aligned}
\left\langle \mathcal{O}_{5} \right\rangle &= -\frac{15 \mathcal{C}s}{2}E
_{3}u + \mathcolor{Black}{\frac{5 \mathcal{C}ps}{2} \left( 2 E _{5} - 3 E
_{3}' \right)} \\
        &\quad - \frac{5 \mathcal{C}s ^{3}}{2} E _{5} -\frac{5 \mathcal{C}^{2}s}{8} \left( E _{5}-2 E _{3}' \right) \ ,
\end{aligned}
\end{equation}

Once again, these results are in good agreement with earlier results
\cite{Beccaria:2017rfz}. The above results are also in agreement with
independent computations of the ring relations that do not take into
account gravitational corrections. For example, \cref{eq:su2-o4} is in
perfect agreement with \cite[eq.~(4.30)]{Fucito:2005wc} when all
gravitational corrections are neglected.

It is also not difficult to push these computations to higher chiral
traces. For example:
\begin{equation}
\begin{aligned}
  \left\langle \mathcal{O}_{6} \right\rangle &= \frac{u ^{3}}{4} + \frac{\mathcal{C}u ^{2}}{8} \left( E _{2}-1 \right)-\frac{3p}{2} u u' - \frac{\mathcal{C}u'}{4} \left( E _{2}-1 \right) + p ^{2} u''  \\
  & \quad u \left[ \mathcolor{Black}{\frac{\mathcal{C}p}{192} \left( 11 E
  _{2}^{2}+ E _{4}-12 \right)} - \frac{\mathcal{C}s ^{2}}{16} \left( E
  _{4}-1 \right) - \frac{\mathcal{C}^{2}}{64} \left( E _{2}-1 \right)
  \right] \\
  &\quad + \mathcolor{Black}{\frac{\mathcal{C}p ^{2}}{40320} \left( 245 E
  _{2}^{3}+21 E _{2} E _{4}-26 E _{6}-240 \right)} \\
  & \quad + \mathcolor{Black}{\frac{\mathcal{C}ps ^{2}}{1120} \left( -21 E
  _{2} E _{4} + E _{6}+20 \right)} \\
  & \quad \mathcolor{Black}{+ \frac{\mathcal{C}^{2}p}{120960} \left( - 140
  E _{2}^{3} -84 E _{2} E _{4} -105 E _{2}^{2} +105 E _{4} + 44 E _{6} + 180
  \right)} \\
  & \quad + \frac{\mathcal{C}s ^{4}}{168} \left( E _{6}-1 \right) + \frac{\mathcal{C}^{2} s ^{2}}{2240} \left( 25200 E _{3}^{2} + 7 E _{2} E _{4}-2 E _{6} - 5 \right) \\
  & \quad + \frac{\mathcal{C}^{3}}{483840} \left( -140 E _{2}^{3} +84 E _{2} E _{4} -24 E _{6} +525 E _{2}^{2} -420 E _{2} + 80 \right) \ .
\end{aligned}
\end{equation}
Similar expressions for higher traces are easily derived and
resummed. We do not reproduce the expressions here since they are
rather long and not proportionately enlightening.

\subsection*{\boldmath $ \mathrm{SU}(3) $}
The above computations are easily generalised for the case of higher
traces (at fixed $ N $) and also higher-rank gauge groups. For
example, for $ N=3 $ we have:
\begin{equation}
\begin{aligned}
  \left\langle \mathcal{O}_{4} \right\rangle &= \frac{u _{2} ^{2}}{2} - p u _{2}' + \frac{3\mathcal{C}}{48} \left( E _{2}-1 \right) u _{2} \\
        &\quad + \frac{\mathcal{C}p}{480} \left( 5 E _{2}^{2} - 2E _{4} - 3 \right) - \frac{\mathcal{C}s ^{2}}{80} \left( E _{4}-1 \right) \\
        &\quad \quad - \frac{\mathcal{C}^{2}}{2560} \left( 15 E _{2}^{2}-6 E _{4}-10 E _{2}+1 \right) \ ,
\end{aligned}
\end{equation}
\begin{equation}
\begin{aligned}
        \left\langle \mathcal{O}_{5} \right\rangle &= \frac{5}{6} u _{2} u _{3} + -\frac{5p}{3} u _{3}' + \frac{5\mathcal{C}}{32} \left( E _{2}-1 \right) u _{3} -\frac{45 \mathcal{C}s}{8} E _{3} u _{2} \\
        &\quad - \frac{15 \mathcal{C}s ^{3}}{4} E _{5} + \frac{15
        \mathcal{C}ps}{2} \left( E _{5}-2 E _{3}' \right) \\
        &\quad \quad + \frac{15 \mathcal{C}^{2}s}{128} \left( -8 E _{5}+3 E _{2}E _{3} +24 E _{3}' -3 \right)  \ ,
\end{aligned}
\end{equation}
\begin{equation}
\begin{aligned}
          \left\langle \mathcal{O}_{6} \right\rangle &= \frac{u _{2} ^{3}}{4} + \frac{u _{3,3}}{3} - \frac{3p}{2} u _{2} u _{2}' + p ^{2} u _{2}'' + \frac{7 \mathcal{C}}{64} \left( E _{2}-1 \right) u _{2}^{2}  \\
          & \quad - \frac{7 \mathcal{C}p}{32} \left( E _{2}-1 \right) u
          _{2}' + u _{2} \left[ \mathcolor{Black}{\frac{\mathcal{C}p}{96}
          \left( 7 E _{2}^{2}-E _{4}-6 \right)}-\frac{\mathcal{C}s ^{2}}{16}
          \left( E _{4}-1 \right) \right. \\
          & \quad \quad \left. + \frac{\mathcal{C}^{2}}{1024} \left( - 9 E _{2}^{2}+E _{4}- 6 E _{2}+11 \right) \right] -\frac{27 \mathcal{C}s}{2} E _{3} u _{3}  \\
          & \quad \mathcolor{Black}{+ \frac{\mathcal{C}p ^{2}}{5040} \left(
          70 E _{2}^{3}-21 E _{2} E _{4} -4 E _{6}-45 \right)} \\
          & \quad \mathcolor{Black}{- \frac{3\mathcal{C} p s ^{2}}{560}
          \left( 7 E _{2} E _{4}-2 E _{6} -5 \right) } \\
          & \quad \quad \mathcolor{Black}{+ \frac{\mathcal{C}^{2}p}{10752}
          \left( -35 E _{2}^{3}+7 E _{2} E _{4}+4 E _{6}-21 E _{2}^{2}+21 E
          _{4}+24 \right)}\\
          & \quad \mathcolor{Black}{\frac{\mathcal{C}s ^{4}}{112} \left( E
          _{6}-1 \right) -\frac{3\mathcal{C}^{2} s ^{2}}{8960} \left( 45360
          E _{3}^{2} - 21 E _{2} E _{4} + 11 E _{6}+10 \right)} \\
          & \quad \mathcolor{Black}{+ \frac{\mathcal{C}^{3}}{573440} \left(
          -105 E _{2}^{3}+84 E _{2}E _{4}-24 E _{6}\right.} \\
          & \quad \quad \quad \quad \quad \quad \quad \quad
          \mathcolor{Black}{\left. +735 E _{2}^{2}-280 E _{4}-455 E _{2} +45
          \right)} \ .
\end{aligned}
\end{equation}

It is noteworthy that the gravitational corrections to the trace
relations cannot be reconstituted into (quasi-)modular forms of a
definite weight. It follows that any theorems regarding the
finite-dimensionality of the space of weight-$ k $ quasimodular forms
cannot be brought to bear on the problem at hand. The only solution in
this case is to pursue the computation of trace relations to higher
and higher instanton number, a task that is \emph{significantly}
easier using the regularity of $ qq $-characters.

In concluding this section, we remark that the relative ease with
which these results are arrived at should be abundantly clear. If one
had to arrive at the same results using equivariant localisation of an
$ \mathrm{SU}(N) $ gauge theory, one would have to perform $ N-1 $
series inversions to express the classical vevs $ a _{u} $ in terms of
the gauge-invariant Coulomb moduli and their derivatives. One would
then plug these inversions (or mirror maps) into the expression for
the chiral correlator computed via localisation to derive the
corresponding (deformed) ring relations. In the $ qq $-character
framework, none of this is required, since we only work with gauge
invariant Coulomb moduli.

\section{Discussion}
\label{sec:discussion}
We have verified that requiring the regularity of fundamental
$ qq $-character yields the deformed trace relations in mass-deformed
$ \mathcal{N}=4 $ super Yang-Mills theories with $ \mathrm{SU}(N) $
gauge groups and in this paper have reported on the results of these
studies for the case of low $ N $. That the above computations can be
easily extended to higher rank is abundantly clear, and we have only
refrained from reporting these results for even higher rank in the interest
of brevity. We conclude this short note with a few observations and
directions for future work.

\subsection*{Good Operators}
We have seen that the differential ring structure acquired by chiral
operators in the presence of an $ \Omega $-background arises
principally because of \cref{eq:u2-shift}, which in turn is a simple
consequence of Matone's relation in the presence of gravitational
couplings \cite{Flume:2004rp}. It is easy to see, then, that one can
systematically make nonperturbatively exact redefinitions of the
chiral operators $ \mathcal{O}_{n} $ in the Nekrasov-Shatashvili limit
$ (p \rightarrow 0, s \neq 0) $, in close analogy with what can be
done in the undeformed theory \cite{Ashok:2016ewb}, so that the new
operators satisfy the classical chiral ring relations. The exact form
of the redefinitions can be read off from the appropriate expressions
in \Cref{sec:results}.

\subsection*{Bloch-Okounkov}
Given a function $ g(\lambda ) $ on the set of integer partitions, its
Bloch-Okounkov $ q $-bracket \cite{BlochOkounkov2000} is defined as
\begin{equation}
\left\langle g \right\rangle _{q} = \frac{\sum_{\lambda}^{} g(\lambda ) \, q ^{\left\vert \lambda  \right\vert }}{\sum_{\lambda }^{} q ^{\left\vert \lambda  \right\vert }} \ .
\end{equation}
This may be thought of as a statistical average of $ g(\lambda ) $ in
a canonical ensemble at inverse temperature
$ \beta = 2 \pi \tau / i $, for a system whose every energy state
$ \left\vert \lambda \right\vert $ has
$ p(\left\vert \lambda \right\vert ) $ microstates associated to
it.\footnote{Here, $ p(n) $ is the number-theoretic partition
  function, i.e.~it counts the number of partitions of $ n $.}
The denominator, then, is simply the thermal partition function of
this statistical system. Now, for $ k \in \mathbb{N} $ and associated
to a partition $ \lambda $ as described above, define
\begin{equation}
S _{2k}(\lambda ) = \sum_{j}^{} \lambda _{j}^{2k-1} \ .
\end{equation}
(The case $ k=1 $ is simply
$ S _{2}(\lambda ) = \left\vert \lambda \right\vert $, i.e.~the number
of boxes in the Young diagram.) A beautiful result of
\cite{Zagier2016} finds that the Bloch-Okounkov $ q $-bracket of these
functions $ S _{2k}(\lambda ) $ are
\begin{equation}
\left\langle S _{2k} \right\rangle _{q} = \frac{B _{2k}}{4k} \left( 1 - E _{2k}(\tau ) \right) \ ,
\end{equation}
i.e.~a quasimodular form of indefinite weight. Further, by relaxing
the above constraint on $ k $ and allowing it to take half-integer
values, we find precisely the ``odd-weight'' Eisenstein series. Note
that these kinds of terms appear repeatedly as coefficients of
gravitational corrections to the trace relations we encountered in
\Cref{sec:results}.  Perhaps, since the fundamental $ qq $-character
in \cref{eq:fund-qq-character-1} can be written (up to normalisation
by the partition function of this statistical system) as a
Bloch-Okounkov $ q $-bracket, it is not so surprising that the
gravitational corrections often admit the same representation. Whether
this observation will help circumvent the need for first studying
theories order-by-order in the instanton expansion and then resumming
observables to establish consistency with considerations of S-duality
remains to be seen.

\subsection*{KdV Charges}
As we mentioned in the introduction to this note, the duality between
four-dimensional gauge theories and two-dimensional conformal field
theories implies, among other things, a correspondence between chiral
operators in the former and the so-called quantum KdV charges in the
latter. The precise nature of this map has not yet been uncovered and
remains an interesting open problem. The deformed chiral ring
relations we have found in this paper will serve as a strong
consistency check on any such correspondence, although as observed by
\cite{Beccaria:2017rfz}, it is insufficient to fix it
completely. There has nevertheless been considerable progress in
understanding and constructing quantum KdV charges
\cite{Bazhanov:1994ft,Maloney:2018yrz,Maloney:2018hdg,Dymarsky:2018iwx,Dymarsky:2018lhf}
and this work has recently been extended to the case of quantum
Boussinesq charges associated to the $ \mathcal{W}_{3} $ algebra
\cite{Ashok:2024zmw}. We hope to return to this question in the near
future.

\appendix

\section{Fourier Expansions}
\label{sec:fourier-expansions}
Let $ \sigma _{k}(n) $ be the divisor function, defined as
\begin{equation}
\sigma _{k}(n) = \sum_{d \vert n}^{} d ^{k} \ ,
\end{equation}
i.e.~the sum of all divisors $ d $ of $ n $, each raised to the power
$ k $. Using these divisor functions, for $ k \in \mathbb{N} $ we
define the Eisenstein series
\begin{equation}
E _{2k}(\tau) = 1 - \frac{4k}{B _{2k}} \sum_{n=1}^{\infty } \sigma _{2k-1}(n) \, q ^{n} \ ,
\end{equation}
where $ B _{n} $ is the $ n $th Bernoulli number. The Eisenstein
series so defined are said to have weight $ 2k $ because under a
transformation
\begin{equation}
S: \tau \mapsto - \frac{1}{\tau } \ ,
\end{equation}
we find that for $ k \geq 2 $ the Eisenstein series transform
covariantly as
\begin{equation}
E _{2k}\left( -1/\tau \right) = \tau ^{2k} \, E _{2k}(\tau ) \ .
\end{equation}
The Eisenstein series are examples of modular forms. The case $ k=1 $
is anomalous, in that under an $ S $-transformation we find
\begin{equation}
E _{2}\left( - 1/\tau  \right) = \tau ^{2} E _{2}(\tau ) + \frac{6 \tau }{i \pi }  \ ,
\end{equation}
and for this reason it is often referred to as a quasimodular form.

We have also used the ``odd weight'' Eisenstein series
\begin{equation}
E _{2k+1}(\tau ) = \sum_{n=1}^{\infty } \sigma _{2k}(n) \, q ^{n} \ .
\end{equation}
frequently in our resummations. These ``odd weight'' Eisenstein series
are not modular \cite{Shimomura}, but satisfy a property called
holomorphic quantum modularity \cite{Zagier_2020}.

\bibliographystyle{amsplain}
\bibliography{refs}
\end{document}